\documentclass[english,aps,floats,twocolumn,showpacs,prl,amsfonts]{revtex4}

\usepackage{pslatex,amssymb,amsmath,amsfonts,graphicx,epsfig}
\usepackage[T1]{fontenc}
\usepackage[latin1]{inputenc}

{
{
\newcommand{\be}{\begin{equation}}
\newcommand{\ee}{\end{equation}}
\newcommand{\bea}{\begin{eqnarray}}
\newcommand{\eea}{\end{eqnarray}}

\begin{document}

\title{Inflation in minimal left-right symmetric model with spontaneous
$D$-parity breaking}

\author{Jinn-Ouk Gong}
\email{jgong@hep.wisc.edu}%
\affiliation{Harish-Chandra Research Institute, Chhatnag Road, Jhunsi, Allahabad,
211 019, India\footnote{Present address: Department of Physics, University of
Wisconsin-Madison, 1150 University Avenue, Madison, WI 53706-1390, USA}}

\author{Narendra Sahu}
\email{n.sahu@lancaster.ac.uk}%
\affiliation{Theory Division, Physical Research Laboratory, Navrangpura, Ahmedabad,
380 009, India\footnote{Present address: Cosmology and Astroparticle Physics Group,
University of Lancaster, Lancaster LA1 4YB, UK}}

\begin{abstract}
We present a simplest inflationary scenario in the minimal left-right
symmetric model with spontaneous $D$-parity breaking, which is a
well motivated particle physics model for neutrino masses.
This leads us to connect the observed anisotropies in the cosmic microwave
background to the sub-eV neutrino masses. The baryon asymmetry via the
leptogenesis route is also discussed briefly.

\end{abstract}


\maketitle

It is now widely believed that the Universe has gone through a period of
inflation~\cite{inf} at the earliest moment of its history. Inflation is
required to explain finely tuned initial conditions of the standard hot big
bang cosmology, as well as to solve many cosmological problems such as
homogeneity, isotropy and flatness of the observable Universe. Moreover, it
is predicted that during inflation primordial density perturbations, necessary
for large scale structure in the Universe and the temperature fluctuations in
the cosmic microwave background (CMB), are generated from quantum fluctuations.
The mechanism of inflation is now a well established subject~\cite{books}, and
recent observations of the galaxy distribution and the CMB are in strong favor
of inflation~\cite{obs}.

It is, however, still unclear how to build a realistic and sensible
scenario of inflation in particle physics. Because of the extremely
high energy scale of the early universe where inflation takes place, it
is usually believed that the particle physics models, invoked as a plausible
framework to implement inflation, would possess larger symmetries than the
standard model (SM) of particle physics. Supersymmetry (SUSY)
and grand unified theories (GUTs) are such popular extensions of the
SM~\cite{susy-gutreview}.

An attractive extension of the SM is the minimal left-right symmetric
model~\cite{left-right-group} with spontaneous $D$- parity
breaking~\cite{paridaetal.84}. The advantages of considering this model
is that (a) it has a natural explanation for the origin of parity violation
which is preferential under the SM gauge group $SU(3)_C\times SU(2)_L\times
U(1)_Y$, (b) it can be easily embedded in the $SO(10)$ GUT, and (c) $B-L$ is
a gauge symmetry: since $B-L$ is a gauge symmetry of the model, it is not
possible to have any $L$-asymmetry~\cite{fukugita.86} before the left-right
gauge symmetry breaking. A net $L$-asymmetry is produced after the $B-L$ gauge
symmetry breaking phase transition. The $L$-asymmetry is then transferred to
the required baryon asymmetry in the presence of the non-perturbative
electroweak processes which conserve $B-L$ but violate $B+L$.

In this letter we present an inflationary scenario embedded in such an
extension of the SM. We saw that in contrast to the conventional left-right
symmetric model where $D$-parity breaks at ${\cal O}(10^{16})$ GeV or below,
the inflationary scenario in this model demands $D$-parity should be
broken above GUT scale. Therefore, other than the conventional
successes of the inflationary scenario, it naturally explains the
vanishingly small, but non-zero neutrino masses and the observed baryon
asymmetry through the leptogenesis route. We also saw that in the
certain parameter space the observed anisotropies in the cosmic microwave
background radiation is intimately related to the sub-eV neutrino masses.
Thus our model is not only cosmologically relevant, but also favorable for
the observed particle physics phenomenology.

{\it Left-right symmetric model}: We now recapitulate the salient features
of the minimal left-right symmetric model with spontaneous $D$-parity
violation. The gauge group of the model is given by $SU(2)_L\times SU(2)_R
\times U(1)_{B-L} \times P$. At a high scale $(10^{16} \sim 10^{19})$ GeV
the parity is broken by a singlet field $\sigma (1,1,0)$, with the numbers
inside the parentheses being the quantum numbers under the gauge group, and
it leaves the gauge symmetry $SU(2)_L\times SU(2)_R \times U(1)_{B-L}$
intact. At a comparative low scale $SU(2)_R \times U(1)_{B-L}$ gauge
symmetry is broken to $U(1)_Y$ by a triplet scalar $\Delta_R (1,3,2)$.
Through the Majorana Yukawa coupling $\Delta_R$ gives masses to the
right-handed neutrinos which anchor the canonical seesaw
mechanism~\cite{typeI_seesaw} to give small Majorana masses to the left
handed physical neutrinos. The left-right gauge symmetry requires another
triplet $\Delta_L (3,1,2)$ whose vacuum expectation value (VEV)
gives masses to the physical left handed neutrinos through the triplet
seesaw~\cite{typeII_seesaw}. Finally $SU(2)_L\times U(1)_Y$ is broken to
$U(1)_{em}$ by a bidoublet $\Phi (2,2,0)$ which essentially contains two
copies of $SU(2)$ doublets with opposite hypercharge. This gives masses
to all the SM fields. Under the left-right parity the scalars transform as
\begin{equation}
\sigma \leftrightarrow -\sigma, ~~~\Delta_R \leftrightarrow \Delta_L
~~\mathrm{and} ~~\Phi \leftrightarrow \Phi^\dagger\,.
\end{equation}
On the other hand, the fermion doublets $\psi_L^T (2,1,-1)\equiv
(\nu_L, e_L)$ and $\psi_R^T (1,2,-1)\equiv (\nu_R, e_R)$ under the
left-right parity transform as $\psi_L \leftrightarrow \psi_R$.

Since $\sigma$ is a singlet field under the gauge group $SU(2)_L\times
SU(2)_R \times U(1)_{B-L}$ it may dominate the energy density of the
Universe for some duration and hence can play the role of the inflaton
field~\cite{Berezhiani:1995am}. As we will see soon, inflation occurs
while $\sigma$ is slowly rolling on its potential towards the minimum.
As soon as $\sigma$ acquires a VEV parity is broken. Therefore, $\sigma$
plays a dual role in this model. However, it does not affect the gauge
symmetry of the group, since as mentioned above it is a singlet under the
remaining gauge group. A bonus point in this model is that inflation solves the
generic domain wall problem by sweeping them away.

We now write down the potential involving the scalar fields
$\Delta_R$, $\Delta_L$, $\Phi$ and $\sigma$. The relevant potential
for the rest of our discussion is given by
\begin{equation}
\mathbf{V}=\mathbf{V}_{\sigma}+\mathbf{V}_{\Phi}+\mathbf{V}_{\Delta}
+ \mathbf{V}_{\sigma\Delta}+\mathbf{V}_{\sigma\Phi}+
\mathbf{V}_{\Phi\Delta}\,,
\label{potential}
\end{equation}
where
\begin{align}\label{potentialpieces}
\mathbf{V_\sigma} =& -\frac{1}{2}\mu^2 \sigma^2 + \frac{1}{4}\lambda
\sigma^4\nonumber + V_0\,,
\nonumber\\
\mathbf{V}_\Delta =& -\mu _{\Delta}^2\left[ Tr\left( \Delta _L \Delta
_L^{\dagger}\right) + Tr\left( \Delta _R\Delta _R^{\dagger }\right) \right]+ {\rm
quartic\,\,terms}\,,
\nonumber\\
\mathbf{V}_{\sigma \Delta} =& M \sigma \left[Tr(\Delta_R
\Delta_R^\dagger) - Tr(\Delta_L \Delta_L^\dagger)\right]\nonumber\\
&+\gamma \sigma^2 \left[ Tr(\Delta_L \Delta_L^\dagger)
 + Tr(\Delta_R \Delta_R^\dagger)\right]\,,
\nonumber\\
\mathbf{V}_{\Phi \Delta} =& \beta \left[ Tr\left( \widetilde{\Phi }\Delta _R \Phi
^{\dagger }\Delta _L^{\dagger }\right) + Tr\left( \widetilde{\Phi }^{\dagger }\Delta
_L\Phi \Delta _R^{\dagger }\right) \right] + {\rm \cdots}\,,
\end{align}
where $\mu$ and all $\mu_a$, with $a$ denoting $\Delta$, $\Phi$, and
$\widetilde{\Phi}=\tau_2 \Phi^* \tau_2$, are positive. $\mathbf{V}_{\Phi}$
and $\mathbf{V}_{\sigma\Phi}$ are chosen in such a way that $\Phi$ acquires
a VEV and hence breaks the gauge symmetry $SU(2)_L\times U(1)_Y$ down to
$U(1)_{em}$. In $\mathbf{V_\sigma}$, $V_0$ is a constant and properly
chosen so that the minimum of the potential $\mathbf{V}_\sigma$ settles
at zero.

As the Universe expands, the temperature falls so that below the critical
temperature $T_c\equiv \sigma_P$, $\sigma$ acquires a VEV
\begin{equation}
\langle \sigma \rangle \equiv \sigma_P = \frac{\mu}{\sqrt{\lambda}}\,.
\end{equation}
As a result, the effective masses of the triplets $\Delta_L$ and
$\Delta_R$ are given by
\begin{align}
M_{\Delta_R}&=\sqrt{\mu_\Delta^2-(M \sigma_P+\gamma \sigma_P^2)}\,\,,
\nonumber\\
M_{\Delta_L}&=\sqrt{\mu_\Delta^2+(M \sigma_P-\gamma \sigma_P^2)}\,.
\label{triplet_masses}
\end{align}
We now do a fine tuning to set $M_{\Delta_R}^2 > 0$, so that it
acquires a VEV
\be
\langle \Delta_{R}\rangle =
\begin{pmatrix}
  0 & 0 \\
  v_R & 0 \\
\end{pmatrix}\,.
\label{right_vev}
\ee
At a few hundred GeV $\Phi$ and $\widetilde{\Phi}$ will acquire VEVs
\be
\langle \Phi \rangle =
\begin{pmatrix}
  k_1 & 0 \\
  0 & k_2 \\
\end{pmatrix}
~~{\rm and}~~\langle \widetilde{\Phi} \rangle =
\begin{pmatrix}
  k_2 & 0 \\
  0 & k_1 \\
\end{pmatrix}\,.
\label{dirac_vev}
\ee
However, this induces a non-trivial VEV for the
triplet $\Delta_L$ as
\be
\langle \Delta_{L}\rangle =
\begin{pmatrix}
  0 & 0 \\
  v_L & 0 \\
\end{pmatrix}\,.
\label{left_vev}
\ee
This gives masses to neutrinos through type-II seesaw. Therefore, it
is worth checking the order of magnitude of $v_L$. From $\mathbf{V}_\Delta$,
$\mathbf{V}_{\sigma \Delta}$ and $\mathbf{V}_{\Phi \Delta}$ of
Eq.~(\ref{potentialpieces}) we get
\begin{equation}
v_R\frac{\partial \mathbf{V}}{\partial v_L}- v_L
\frac{\partial \mathbf{V}}{\partial v_R} = v_Lv_R[4 M \sigma_P ] +
2 \beta k_1^2(v_R^2-v_L^2) = 0\,.
\label{minimisation}
\end{equation}
Observed phenomenology requires $v_L \ll k_2 < k_1 \ll v_R$. Thus
the above equation gives
\be
v_L \approx \frac{-\beta v^2v_R}{2M\sigma_P}\,,
\label{vev_value}
\ee
where we have used $v=\sqrt{k_1^2+k_2^2}\approx k_1=174$ GeV and $\beta$ is a
coupling constant of $\mathcal{O}(1)$. Notice that in the above equation the
smallness of the VEV of $\Delta_L$ is decided by the parity breaking scale,
but not the $SU(2)_R$ breaking scale~\footnote{If the parity and $SU(2)_R$
are broken at the same scale then the smallness of $v_L$ depends on the
large value of $v_R$ through the seesaw relation $v_L v_R\approx v^2$, which
implies $v_R\approx \left(10^{13} \sim 10^{14}\right)$ GeV to obtain
$v_L\approx \mathcal{O}(1)$ eV.}. So there are no constraints on $v_R$ from
the type-II seesaw point of view.

{\it Inflation by $\sigma$}: As mentioned before, since $\sigma$ is a singlet
its energy density dominates the total energy density of the Universe and
hence is able to drive inflation. From $\mathbf{V_\sigma}$ of
Eq.~(\ref{potentialpieces}) we can see that the choice $V_0 =
\mu^4/(4\lambda)$ sets the minimum of the potential to be zero. We now write
the slow-roll parameters in terms of $V(\sigma)$ as
\begin{equation}\label{SRparameters}
\epsilon  \equiv \frac{M_\mathrm{Pl}^2}{16\pi} \left( \frac{V'}{V} \right)^2 \,
~~{\rm and} ~~~\eta  \equiv \frac{M_\mathrm{Pl}^2}{8\pi} \frac{V''}{V} \, ,
\end{equation}
where $M_\mathrm{Pl} \equiv G^{-1/2} \approx 1.22 \times 10^{19} \,
\mathrm{GeV}$ is the Planck mass and the prime denotes a derivative with
respect to $\sigma$. Inflation ends when the scale factor accelerates no
more, and this happens when $\epsilon_\mathrm{end} = 1$. This gives
\begin{equation}
\sigma_\mathrm{end}^2 \approx \frac{\mu^4}{4\lambda \left( \lambda
M_\mathrm{Pl}^2/(4\pi) + \mu^2 \right)}\,.
\end{equation}
Thus the number of $e$-folds from $\sigma$ to $\sigma_\mathrm{end}$ can
be estimated as
\begin{align}\label{efolds}
N(\sigma) = & -\frac{8\pi}{M_\mathrm{Pl}^2} \int_\sigma^{\sigma_\mathrm{end}}
\frac{V}{V'} d\sigma
\nonumber\\
= & \frac{\pi \mu^2 }{\lambda M_\mathrm{Pl}^2} \log \left[ \frac{\mu^4}{4
\lambda \left( \lambda M_\mathrm{Pl}^2/(4\pi) + \mu^2 \right) \sigma^2}
\right]\nonumber\\
& - \frac{\pi}{M_\mathrm{Pl}^2} \left[ \frac{\mu^4}{4 \lambda
\left( \lambda M_\mathrm{Pl}^2/(4\pi) + \mu^2 \right) \sigma^2} -
\sigma^2 \right]\,,
\end{align}
where we note that the contribution from the second term is much less
than that from the first term. From the observed amplitude of the
density perturbations on the COBE scale \cite{Bunn:1996da}
\begin{equation}
\delta_H = \sqrt{\frac{1}{75\pi^2m_\mathrm{Pl}^6} \frac{V_H^3}{{V_H'}^2}}
\approx 1.91 \times 10^{-5} \, ,
\end{equation}
we can find the corresponding value of $\sigma$ as
\begin{equation}\label{COBEfield}
\sigma_H^2 \approx \frac{8\pi^3\mu^8}{\lambda^3A_H^2 M_\mathrm{Pl}^6} \, ,
\end{equation}
where $A_H \equiv \sqrt{75}\pi \delta_H \approx 5.19 \times 10^{-4}$. Then
we can easily estimate the spectral index at the COBE point as \cite{index}
\begin{equation}\label{index}
n_\mathrm{s} \approx 1 - \frac{\lambda M_\mathrm{Pl}^2}{\pi\mu^2} -
\frac{40\pi^2\mu^4}{\lambda A_H^2 M_\mathrm{Pl}^4}\,.
\end{equation}

\begin{figure*}[t]
\begin{center}
\epsfig{file=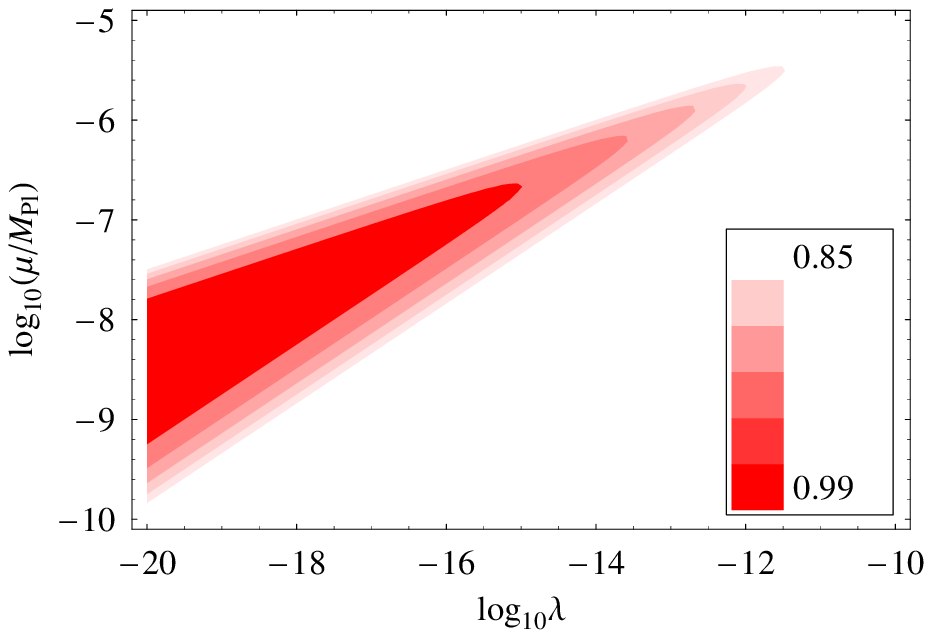, width=0.5\textwidth}%
\epsfig{file=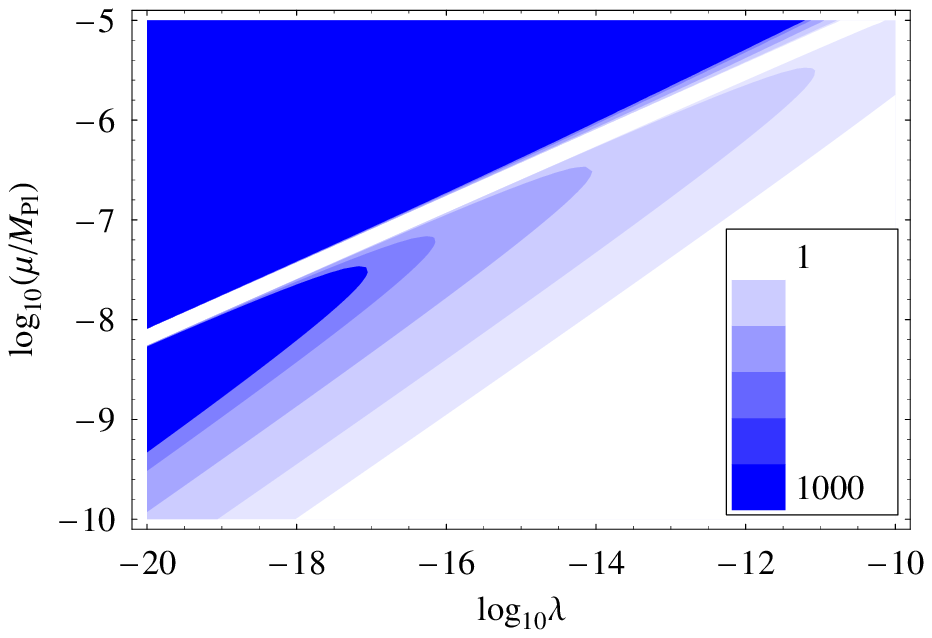, width=0.5\textwidth}
\end{center}
\caption{The contour plots of (left) $n_\mathrm{s}$ and (right) $N_H$. The
horizontal and vertical axes are $\log_{10}\lambda$ and $\log_{10}
\left(\mu/M_\mathrm{Pl}\right)$, respectively, for both graphs. In the contour plot
of $n_\mathrm{s}$, the contours denote 0.99, 0.97, 0.94, 0.90 and 0.85 from the
innermost line. Likewise, we have set 1000, 500, 100, 10 and 1 in the $N_H$ plot.
Note that in the right panel although we have $N_H \gg 1$ in the upper left region,
the values of $\lambda$ and $\mu$ taken from here will place the minimum of
potential far larger than $M_\mathrm{Pl}$ and the form of the effective potential is
apt to an appreciable modification, spoiling all the results we have estimated.
Thus we disregard the values of $\lambda$ and $\mu$ within this region.}%
\label{contourplot}
\end{figure*}

As a sample set of values, let us take $\mu = 2/\pi \times 10^{-6}
M_\mathrm{Pl} \approx 7.77 \times 10^{12} \, \mathrm{GeV}$ and
$\lambda = 4/\pi^4 \times 10^{-12} \approx 4.11 \times 10^{-14}$: this
set gives the minimum of the potential at $\pi M_\mathrm{Pl}$ with an
inflationary energy scale $\mathcal{O} \left( 10^{16} \right) \mathrm{GeV}$.
From Eqs.~(\ref{efolds}), (\ref{COBEfield}) and (\ref{index}), we
obtain\footnote{In fact, there exists some level of uncertainty on from
which value of $\sigma$ inflation begins. Because of uncertainty
principle, quantum fluctuations $\delta\sigma \approx H_i/(2\pi)$ are so
strong near the origin and the classical downhill motion dominates only
when $\sigma^2 \gtrsim \sigma_i^2 \approx 2\pi\mu^8/(3\lambda^3
M_\mathrm{Pl}^6)$. We can find the ratio $\sigma_H^2/\sigma_i^2 \sim 10^8$,
i.e. $\sigma_H$ lies well within the regime of classical evolution of
$\sigma$ and we need not worry about $\sigma_i$.} $N_H \approx 59.0$ and
$n_\mathrm{s} \approx 0.963$. Also, due to the relatively high inflationary
energy scale, we find a tensor-to-scalar ratio $r$ very
close to the observational sensitivity of near future experiments,
$r \approx 0.0163$. In Fig.~\ref{contourplot}, we show the contour plots
of both $n_\mathrm{s}$ and $N_H$ on the $\lambda$-$\mu$ plane.

After the end of inflation, $\sigma$ eventually starts oscillation around its
minimum $\mu/\sqrt{\lambda}$ and decays into light relativistic particles,
reheating the universe to restore the gauge symmetry $SU(2)_L\times
SU(2)_R\times U(1)_{B-L}$ with the reheating temperature being estimated
as \cite{lindebook}
\begin{equation}
T_\mathrm{RH} \sim \mathcal{O}(0.1) \sqrt{\Gamma_\sigma M_\mathrm{Pl}} \, ,
\end{equation}
where we have taken the number of relativistic degrees of freedom to be
$\mathcal{O}\left( 10^2 \sim 10^3 \right)$.

{\it Neutrino masses and the CMB anisotropies}: The relevant Yukawa couplings that
are giving masses to the three generations of leptons are given by
\begin{align}
-\mathcal{L}_\mathrm{Yukawa} =& h_{ij}\overline{\psi}_{iL} \Phi \psi_{jR} +
\tilde {h}_{ij} \overline{\psi}_{iL} \widetilde{\Phi} \psi_{jR} + h.c.
\nonumber\\
& + f_{ij} \left[ \psi_{iR}^T C i \tau_2 \Delta_R \psi_{jR} + (R \leftrightarrow L)
\right] + h.c.
\label{Yukawa}
\end{align}
The discrete left-right symmetry ensures the Majorana Yukawa coupling $f$ to be the
same for both left and right-handed neutrinos. The breaking of the left-right
symmetry down to $U(1)_{em}$ results in the effective mass matrix of the physical
left handed neutrinos to be
\begin{align}
m_\nu & = \frac{-\beta v^2v_R}{2M\sigma_P} f - \frac{v^2} {v_R} h f^{-1} h^T
\nonumber\\
& = m_\nu^{II}+m_\nu^I\,,
\label{neutrino-mass}
\end{align}
where we have used Eq. (\ref{vev_value}) for type-II contribution and neglected
$\mathcal{O}(k_2/k_1)\approx (m_b/m_t)$ terms in the type-I contribution. Assuming
that $h$, $f$ and $\beta$ are $\mathcal{O}(1)$ couplings, the relative magnitude
of $m_\nu^{I}$ and $m_\nu^{II}$ depend on the parameter space of $v_R, M$ and
$\sigma_P$. In the following we assume that type-II term dominates. This is a
viable assumption for $M < v_R^2/\sigma_P$. In what follows we will work in
this regime and then we have
\begin{equation}
\mathcal{H}\equiv m_\nu m_\nu^\dagger \approx \left( \frac{-\beta v^2v_R}{2M\sigma_P} \right)^2
f f^\dagger\,,
\label{type-II-mass}
\end{equation}
where an appropriate choice of $f$ will explain the leptonic mixing. $\mathcal{H}$
can be diagonalised by using the $U_{PMNS}$ matrix and then we will get the solar
and atmospheric mass scales
\begin{align}
\Delta m^2_{\circ} \equiv &\, m_2^2 - m_1^2=\left( \frac{-\beta v^2v_R}{2M\sigma_P} \right)^2
\Delta f_{12}^2 \, , \nonumber\\
\Delta m^2_{atm} \equiv &\, |m_3^2-m_2^2|=\left( \frac{-\beta v^2v_R}{2M\sigma_P} \right)^2
|\Delta f_{23}^2| \, ,
\end{align}
where $\Delta f_{12}^2=f_2^2 -f_1^2$ and $\Delta f_{23}^2=f_3^2-f_2^2$. Using Eq.
(\ref{COBEfield}) in the above equation we get the solar and atmospheric mass scales
to be
\begin{align}
\Delta m_{\circ}^2 =& \left( \frac{-\beta v^2v_R}{2 M M_\mathrm{Pl}} \right)^2
\left( \frac{8 \pi \mu^2}{75 \sigma_H^2}\right)^{1/3} \Delta f_{12}^2
\delta_H^{-2/3} \, , \label{sol_mass_link}\\
\Delta m_{atm}^2 =& \left( \frac{-\beta v^2v_R}{2 M M_\mathrm{Pl}} \right)^2
\left( \frac{8 \pi \mu^2}{75 \sigma_H^2}\right)^{1/3} |\Delta f_{23}^2| \delta_H^{-2/3}\,.
\label{atm_mass_link}
\end{align}
In the above equations $\mu$ can be determined from the precise measurement of
$n_\mathrm{s}$ in the future CMB experiments. Notice that Eqs.~(\ref{sol_mass_link})
and (\ref{atm_mass_link}) give an {\em important relation} between the observed
neutrino mass scales $\Delta m^2_{\circ}$ and $\Delta m^2_{atm}$, and the amplitude
of perturbations on the CMB scale predicted by inflationary scenario in left-right
symmetric models with spontaneous $D$-parity breaking. This is an important prediction
of the theory.

{\it Lepton asymmetry}: Assuming a normal hierarchy in the right-handed neutrino
sector, the decay of the lightest right-handed neutrino can give rise to a net
lepton asymmetry through
\begin{equation}
N_1 \to \left\{
\begin{array}{l}
e^-_{iL} + \phi_1^+
\\
e^+_{iL} + \phi_1^- \, ,
\end{array}
\right.
\end{equation}
where $N_1=\left[\nu_{1_R} +(\nu_{1_R})^c\right]/\sqrt{2}$. The CP
asymmetry in the above decay process is estimated to be
\begin{equation}
\delta_\mathrm{CP} \approx -\frac{1}{8\pi}\left( \frac{f_1}{f_2}\right) \frac{\Im
\left(h^\dagger h \right)_{12}^2}{\left(h^\dagger h\right)_{11}}\,,
\end{equation}
where $f_1$ and $f_2$ are two of the eigenvalues of $f$ matrix, and we have
neglected $\mathcal{O}(k_2/k_1) \approx (m_b/m_t)$ terms. The lepton asymmetry is
then transferred to the required baryon asymmetry through the electroweak sphaleron
processes which conserve $B-L$ but violate $B+L$. A successful baryon asymmetry
requires a lower bound on the mass scale of the lightest right-handed neutrino to be
$M_1\gtrsim 4.8\times 10^8$ GeV~\cite{type-II-bound}.

{\it Conclusions and outlooks}: We have seen that within the left-right symmetric
model inflation is possible only if the left-right parity and $SU(2)_R$ gauge
symmetry are broken at different scales. In particular, the left-right parity is
broken at $\mathcal{O}(M_\mathrm{Pl})$, while leaving $SU(2)_R$ gauge symmetry preserved
until $\mathcal{O}(10^{14})$ GeV or so. As a standard routine, after inflation the
Universe is reheated to restore the left-right gauge symmetry $SU(2)_L \times
SU(2)_R \times U(1)_{B-L}$. As a result a net baryon asymmetry, required for
successful big bang nucleosynthesis, could be generated through the leptogenesis
route. An important prediction in this model is that the neutrino masses are
connected to the anisotropies in the CMB predicted by inflation. We conjecture
that this can be implemented in the $SO(10)$ model which, at present, is the most
favorable scenario for neutrino masses and mixings. Since $\{210\}$ field contains
a $SU(4)_C\times SU(2)_L \times SU(2)_R$ singlet it can play the role of
$\sigma$ as in the present case. This is under consideration and will be reported
separately.

{\it Acknowledgment}: It is our pleasure to thank Subhendra Mohanty, Rabindra N.
Mohapatra, M. K. Parida, Raghavan Rangarajan and Utpal Sarkar for useful comments
and discussions. We also thank Konstantinos Dimopoulos and David Lyth for their
comments and suggestions.

\end{document}